# Modeling Review Spam Using Temporal Patterns and Co-bursting Behaviors

Huayi Li[†], Geli Fei[†], Shuai Wang[†], Bing Liu[†], Weixiang Shao[†], Arjun Mukherjee[‡], Jidong Shao[§]

[†]Department of Computer Science
University of Illinois at Chicago, IL, USA
*{hli47, gfei2, swang207, liub, wshao4}@uic.edu*
[‡]Department of Computer Science
University of Houston, TX, USA
*arjun@cs.uh.edu*
[§]Dianping Inc., Shanghai, China
*jidong.shao@dianping.com*

## ABSTRACT

Online reviews play a crucial role in helping consumers evaluate and compare products and services. However, review hosting sites are often targeted by opinion spamming. In recent years, many such sites have put a great deal of effort in building effective review filtering systems to detect fake reviews and to block malicious accounts. Thus, fraudsters or spammers now turn to compromise, purchase or even raise reputable accounts to write fake reviews. Based on the analysis of a real-life dataset from a review hosting site (dianping.com), we discovered that reviewers' posting rates are bimodal and the transitions between different states can be utilized to differentiate spammers from genuine reviewers. Inspired by these findings, we propose a two-mode Labeled Hidden Markov Model to detect spammers. Experimental results show that our model significantly outperforms supervised learning using linguistic and behavioral features in identifying spammers. Furthermore, we found that when a product has a burst of reviews, many spammers are likely to be actively involved in writing reviews to the product as well as to many other products. We then propose a novel co-bursting network for detecting spammer groups. The co-bursting network enables us to produce more accurate spammer groups than the current state-of-the-art reviewer-product (co-reviewing) network.

## 1. INTRODUCTION

Opinions in reviews are increasingly used by individuals and organizations for making purchase decisions. Positive opinions often mean profits and fames for businesses and individuals, which unfortunately give strong incentives for fraudsters to promote or to discredit some target products or services. Such activities are called opinion spamming [13]. Unlike web spam, opinion spam involves complex user dynamics. Several researchers have studied this problem [7, 14, 15, 19, 20]. Many review hosting companies such as Yelp and Dianping have also built their own filtering systems to identify and to remove (or hide) fake reviews from their product pages. These systems help alleviate the negative influence of fake reviews and greatly increase the cost of spamming. Fraudsters are thus forced to promote or demote products using accounts with high credibility because reviews from such accounts are likely to survive and to be trusted by consumers. Some of those accounts are compromised or bribed by the fraudsters, but the cost is high. Therefore, spammers began to raise their own accounts by behaving like genuine reviewers for a period of time and then use these "trustworthy" accounts to launch spam campaigns. Such accounts are usually called raised accounts (like raising kids) which behave normally simply to gain credibility in some period of time but are later used to write fake reviews that may be ranked higher by review systems to collectively launch spam campaigns.

Our work is motivated by this observation from the review hosting site dianping.com. Using their real-life filtered (fake or spam) and unfiltered (genuine) reviews, we discovered that both spammers and non-spammers exhibit bimodal temporal posting patterns but their patterns differ greatly. Based on this bimodal discovery, we propose to use two-mode Hidden Markov Model to capture the bimodal behaviors. The reviews of a user in the order of posting time form a chain. Hidden states of a user at each time-stamp can be either *active* or *inactive*. Users in active/inactive state post reviews in fast/slow rate respectively. We further extend it to a Labeled Hidden Markov Model so as to estimate the likelihood of each user being a spammer.

We demonstrate the effectiveness of our model by applying it to a large real-life dataset from Dianping.com. Our contributions in this paper are as follows:

1. We discovered the disparate bimodal distributions of review spammers and non-spammers. To the best of our knowledge, this is the first such discovery. Labeled Hidden Markov Model (LHMM) is proposed to detect spammers by modeling their temporal posting behaviors. Note that our work is not to reverse-engineer Dianping's filter which consists of over 100 algorithms with numerous manually designed rules using a large number of signals. Our principled approach requires only the time-stamps of reviews, and it is easily applicable to many other social media platforms. In return, our discovery and algorithm can improve commercial filters too.

2. This work is the first to discover the disparities between active/inactive states transitions between spammers and non-spammers. Such behaviors are encoded in Labeled HMM to help infer users' class labels.

3. Compared with other available academic datasets which were crawled from review host sites, our dataset is shared by Dianping.com directly with their commercial filter's output. Due to its large size and accurate labels, for the first time, we are able to compare the proposed model with state-of-art methods in large scale.

4. Besides detecting individual spammers, the model's hidden states can also be used to find spammer groups who work together on spam campaign. A co-bursting network to measure the collusion of reviewers is proposed. Comparison of clustering results indicates that our network is more effective in detecting spammer groups than the review-product network used in the existing work [24, 34].

## 2. RELATED WORK

### 2.1 Academic Datasets

Jindal and Liu [13] released the first opinion spam dataset[1] crawled from Amazon in 2008. However, they simply treated duplicate and near-duplicate Amazon product reviews as fake reviews. There certainly exist fake reviews that are not identical or similar to any other reviews. Ott et al. [26] used Amazon Mechanical Turk (AMT) to crowdsource anonymous online workers to write fake hotel reviews. The review dataset[2] that they compiled had only 1,600 reviews which are too small to support reliable statistical analysis. Furthermore, the motivations and the psychological states of mind of hired Turkers and the professional spammers in the real world are quite different. As pointed out in [25], these crowdsourced reviews are not similar to real-life fake ones in review hosting sites and they also do not have reviewer behavior information.

Mukherjee and other researchers [17, 25] reported the analysis of commercial filter based on reviews of a small number of hotels and restaurants in Chicago that they crawled[3]. In the product/business page of yelp, on the top there are a list of recommended reviews which are considered as authentic but at the bottom there is a link to a list of reviews that are currently not recommended. Mukherjee et al. assume those down-weighted reviews are fake. There are two problems with this approach: First, low quality reviews or short reviews are generally hidden by Yelp but they do not necessarily represent the fake reviews. Secondly, Yelp has an anti-scraping technique which prevents them from scraping all reviews of the hotels. Due to the difficulty of crawling and Yelp's rate limit, they only obtained a small set of (about 66,000) reviews. Those above-mentioned academic review datasets all suffer from two critical problems, i.e. (a) they under-represent the reviews of hotels/restaurants or users as they did not scape all the reviews for each hotel/restaurant nor each reviewer (see section 5); (b) they rely on pseudo labels which are inferred by the authors' intuition rather then the true output from commercial filters or ground truth. The review datasets in our work is shared by Dianping. Not only it is large and representative, but also each review is tagged with their commercial filters in production. For the first time, we are able to study the essence of opinion spam and compare our proposed approach with existing works.

### 2.2 Bursty Reviews

Bursty reviews have been studied recently by several researchers. Fei et al. [6] studied the review time-series for individual products. They assume reviewers in a review burst of a product are typically related in the sense that spammers tend to work with other spammers and genuine reviewers with other genuine reviewers. They thus applied Loopy Belief Propagation to rank spammers using heuristic propagation parameters. Similarly, Xie et al. [33] analyzed multiple time-series of a single retailer including daily number of reviews, average rating, and ratio of singleton reviews. Their end task is to find the time intervals in which a spam attack happens to a retailer. However our work focuses on studying individual reviewers' reviews time-series and its goal is to identify individual spammers and spammer groups. [30] explored temporal dynamics of spam in Yelp such as buffered and reduced spamming rates but does not model inter arrival times. Other researchers applied various Bayesian approaches to detect anomalies in rating time-series [9, 10, 12]. However, our model only requires the timing of each review and the byproduct of our model can be used to detect spammer groups effectively. Incorporating rating signals to our model will be part of our future work.

### 2.3 Classification and Ranking

To detect review spam, Ott et al. [26] built supervised learning models using unigrams and bigrams and Mukherjee et al. [25] added many behavioral features to improve it. Similarly related is the task of psycholinguistic deception detection which detects lies [23], computer-mediated deception in role-playing games [36] and so on. Besides, with only a small portion of labeled reviews, researchers pointed out that using Positive-Unlabeled Learning (or PU learning) [11, 16, 18, 29] outperforms traditional supervised learning. Since PU learning is not the focus of this work, we treat filtered reviews as positive and unfiltered reviews as negative. In the past few years, many researchers incorporated network relations into opinion spam detection. Most of them constructed a heterogeneous network of users/reviews and products. Some of them employed HITS-like ranking algorithm [32], some applied Loopy Belief Propagation [1, 6, 28] and others utilized collective classification [16, 34]. In this work, we propose to build a network using co-bursting relations and it is shown to be more effective in capturing the intricate relations between spammers especially for raised accounts. Our work can largely benefit these above works with co-bursting.

### 2.4 Spammer Group Detection

The second task of our paper is to detect collusive spammers who work in groups. Although recent progress has been made to uncover such spam groups [4, 24, 34, 35], evaluation of spammer groups was troublesome as the ground truth of large-scale dataset was sadly unavailable. Thanks to the labels from Dianping's system, for the first time, evaluation on the clustering results of our co-burst networking against co-reviewing network is made possible. In section 4, we will discuss effectiveness of our proposed clustering framework for spammer group detection.

## 3. MODELING REVIEWERS' ACTIVITIES

### 3.1 Motivation

One of the simplest models for user temporal activity modeling is the Poisson Process which is a process where events occur continuously and independently at a constant average rate. In probability theory and statistics, the exponential distribution is used to describe the length of time intervals between two events in a Poisson Process. The Probability Density Function (PDF) of an exponential distribution is as (1), where $\lambda$ is the rate parameter. So one can see that as $\lambda$ gets larger, events in the process that we are waiting for to happen tends to happen more quickly, hence the name of rate parameter. A very useful property of exponential distribution is that the mean or expected value of an exponentially distributed random variable $X$ with rate parameter $\lambda$ is exactly the reciprocal of $\lambda$ (2).

---

[1] http://liu.cs.uic.edu/download/data/

[2] http://myleott.com/op_spam/

[3] http://liu.cs.uic.edu/download/yelp_filter/

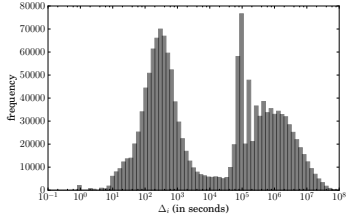

Figure 1: Bimodal distribution of inter-arrival time $\Delta_i$

$$f(x;\lambda) = \begin{cases} \lambda e^{-\lambda x} & x \geqslant 0 \\ 0 & x < 0 \end{cases} \quad (1)$$

$$E(X) = \frac{1}{\lambda} \quad (2)$$

Let $t_i$, $i = 0, 1, \ldots, T$ denote the time-stamps of a user's reviews over a time period of interest. The inter-review duration or inter-arrival time between two adjacent reviews is denoted by $\Delta_i$ and by our assumption $\Delta_i$ is drawn from an exponential distribution with rate parameter $\lambda$.

$$\Delta_i \triangleq t_i - t_{i-1} \quad (3)$$

$$\Delta_i \sim Exp(\lambda) \quad (4)$$

We aggregate all the inter-arrival times between users' adjacent reviews and plot the histogram in Figure 1 to see the distribution in our data (the details of the data are given in the experiment section). Interestingly, the distribution shows two distinct peaks at about 12 minutes and 24 days (we will explain it later in section 3.3). As the Poisson distribution in this setting would typically model the spread of reviews in the next time step around a fixed average, this violates the bimodal distribution of inter-arrival time $\Delta_i$. We then can argue that using homogeneous Poisson Process is insufficient. To solve the problem, we follow the convention of [21] and propose a two-mode HMM to model $\Delta_i$.

### 3.2 Two-mode Hidden Markov Model

A HMM is a model with a sequence of hidden states where one has only observed signals emitted from the hidden states. In the context of our problem, the hidden state $Q_i$ represents *active* or *inactive* mode/state of reviewers and observed signals are the continuous variables $\Delta_i$. $\Delta_i$ between time-stamp $t_{i-1}$ and $t_i$ may follow different exponential distribution depending on $Q_i$. The binary categorization of modes is based on the bimodal distribution shown in Figure 1. Reviews from active mode are written in a *fast* rate while reviews in inactive mode are in a *slow* rate. Both rates are estimated from users' review posting time and they correspond to the two modes/states. We now introduce the hidden states and properties of HMM.

**Hidden States**: We assume that a hidden state variable $Q_i$ takes one of the two possible values $\{0, 1\}$. $Q_i = 0$ denotes that the user is in inactive mode between time-stamp $t_{i-1}$ and $t_i$ while $Q_i = 1$ denotes that the user is in active mode. Our defined model is a first-order Markovian model which assumes $Q_i$ depends only on $Q_{i-1}$ and is independent of previous hidden states $Q_1, Q_2, \ldots, Q_{i-2}$. This approximation is proven reasonable in a great number of applications because it captures short-term memory of human behaviors. Specifically, in our problem, we find strong correlations between consecutive time intervals $\Delta_{i-1}$ and $\Delta_i$. Users in active modes tend to be active and users in inactive modes are more likely to stay inactive. The state transition probability matrix $\mathbb{A}$ is given in (5) where $a_{kj} = P(Q_i = j | Q_{i-1} = k)$, $k, j \in \{0, 1\}$. The initial state probability is a vector $\pi$ and $\pi_j = P(Q_1 = j)$.

$$\mathbb{A} = \{a_{kj}\} = \begin{bmatrix} 1 - \beta_{0,1} & \beta_{0,1} \\ \beta_{1,0} & 1 - \beta_{1,0} \end{bmatrix} \quad (5)$$

**Observation Density**: Since the state variable is unobserved, we can only see the emitted time intervals between two consecutive reviews. In the two-mode HMM, $\Delta_i$'s can be either sampled from fast rate point process when $Q_i = 1$ or slow rate point process when $Q_i = 0$. The two different modes correspond to exponential distributions with rate parameters $\lambda_0$ and $\lambda_1$.

$$\Delta_i \sim \begin{cases} Exp(\lambda_0), & Q_i = 0 \\ Exp(\lambda_1), & Q_i = 1 \end{cases} \quad (6)$$

We now use (6) for drawing $\Delta_i$ with respect to $Q_i$, for $i \in [1, 2, \ldots, T]$. The emission probability distribution is denoted by $\mathbb{B} = \{b_j(\Delta)\}$ and $b_j(\Delta) = f(\Delta; \lambda_j) = \lambda_j e^{-\lambda_j \Delta}$ is the probability of observing some $\Delta$ at state $j$, where $j \in \{0, 1\}$.

Now we can formulate the joint probability of the observations $\Delta_{1:T}$ and hidden states $Q_{1:T}$ as follows.

$$\begin{aligned} P(Q_{1:T}, \Delta_{1:T}) \\ = P(Q_1, Q_2, \Delta_2, \ldots, Q_T, \Delta_T) \\ = P(Q_1) \prod_{i=2}^{T} P(\Delta_i | Q_i) \prod_{i=2}^{T} P(Q_i | Q_{i-1}) \end{aligned} \quad (7)$$

One of the three basic problems for HMM is called the decoding problem which aims to estimate the most likely state sequence in the model given the observations (8). Identifying the hidden states helps to better understand spammers and their collusive behaviors.

$$\begin{aligned} Q_{1:T}^* &= \underset{Q_{1:T}}{\mathrm{argmax}}\, P(Q_{1:T} | \Delta_{1:T}) \\ &= \underset{Q_{1:T}}{\mathrm{argmax}}\, P(Q_{1:T}, \Delta_{1:T}) \end{aligned} \quad (8)$$

A naive approach to examine all possible state assignments has a running time $O(T \cdot 2^T)$ because there are totally $2^T$ possibles combinations and for each such combination it requires $O(T)$ to calculate the product of probabilities. Fortunately, we can employ an efficient dynamic programming algorithm named Viterbi [8] to reduce the time complexity to $O(T \cdot 2^2)$ or simply $O(T)$. Let's define a vector

$$\delta_i(j) = \max_{Q_{1:i-1}} P(Q_{1:i-1}, Q_i = j, \Delta_{1:T}) \quad (9)$$

to store the maximum joint probability along a single path from $Q_1$ to $Q_{i-1}$ when current assignment is $Q_i = j$. On initialization, we set $\delta_1(j) = \pi_j b_j(\Delta_1)$ for $j \in \{0, 1\}$. Then we iteratively calculate the $\delta_i(j)$ using (10) and finally the last state $Q_T^*$ of the most likely state sequence is the one that maximizes (11). Starting from the last state, the sequence of most likely state sequences can be back-tracked through (12).

$$\delta_i(j) = b_j(\Delta_i) \max_{k \in \{0,1\}} \big(\delta_{i-1}(k) a_{kj}\big),\ 2 \leqslant i \leqslant T,\ j \in \{0, 1\} \quad (10)$$

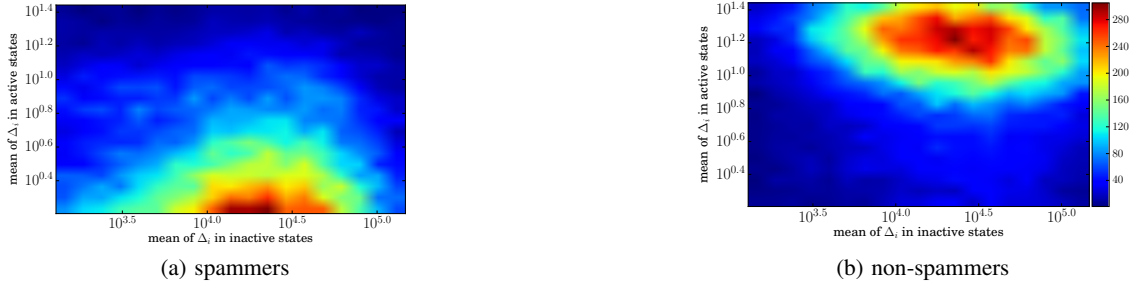

(a) spammers  (b) non-spammers

**Figure 2: Heatmap of inter-arrival time (in seconds) for spammers and non-spammers. Each dot is a user represented by a pair of his mean inter-arrival time in two states ($\mu_0$, $\mu_1$). x-axis is for $\mu_0$ in inactive state, y-axis for is $\mu_1$ in active state.**

$$Q_T^* = \underset{j\in\{0,1\}}{\operatorname{argmax}} \delta_T(j) \quad (11)$$

$$Q_{i-1}^* = \underset{j\in\{0,1\}}{\operatorname{argmax}} \delta_i(j)\, a_{jQ_i^*}, \quad 2 \leqslant i \leqslant T \quad (12)$$

Identifying the state sequence for each reviewer is useful in the sense that reviews from active and inactive states have different impact on calibrating spammers' behaviors. We will show in section 4 that spammers tend to collaborate in active states.

### 3.3 Labeled Hidden Markov Model

Two-mode Hidden Markov Model is very useful for modeling the temporal behavior of reviewers and can be used to predict the timing of the next review. However, it has less predictive power in measuring spammers. Given that our data has commercial level spam labels produced by Dianping's filter, we are able to analyze and compare the dynamics of spammers and non-spammers. Similar to Figure 1, we plot the histogram of aggregated inter-arrival times between spam reviews and their previous reviews (red) as well as between non-spam reviews and their previous reviews (blue) in Figure 3. Note again that x-axis is in log scale. We make the following important observations:

• Bimodal distribution for both classes: Both spam reviews and non-spam reviews show a bimodal distribution. The center of the two humps are far apart from each other indicating distinct two-mode states of review writing patterns. Note that we use the log scale for x-axis. For spam reviews, active states may be the result of aggressive spam activities from a group of spammers active in collusion. For non-spam reviews, active states are likely to happen with events such as sales promotions or holiday celebrations.

• Distinct distributions for active and inactive modes: Since the x-axis of the plot is in log-scale. We can see the histogram for non-spam reviews has much longer tails than spam reviews. This means that a lot of non-spam reviews are written in inactive mode. Besides, there are much more spam reviews in active mode especially less than 100 seconds.

• Disparity of mean of time intervals: For both classes, we simply run Kmeans ($k = 2$) algorithm on the time intervals (log-scale) and compute the mean of inter-arrival times. We find that for both active and inactive states, the mean of time intervals of non-spam reviews are about two to three times longer than that of spam reviews showing a rather normal reviewing activity as opposed to spam reviews that tend to be bursty [6].

Furthermore, we find great difference of user level posting patterns in active and inactive states. We run two-mode HMM on spammers and non-spammers in Dianping's data separately. For each spammer or non-spammer, we calculate the mean of the inter-arrival time for active and inactive mode. Finally, we plot the heatmap (Figure 2) for the two user classes using the pair of mean time intervals. It is interesting to find that spammers and non-spammers distribute very differently. Roughly for $\mu_0$ of inactive states between $[10^4, 10^{4.5}]$, which reflects long activity pauses, $\mu_1$'s of non-spammers are higher than those of spammers by an order of magnitude showing a rather natural behavior of reviewing.

In addition to the disparity of the emission probability one can see from above, we also find interestingly different transition patterns between two states for both spammers and non-spammers. For each of the two user classes, we aggregate the consecutive time intervals between user's reviews and visualize the distribution of all pairs of previous time interval $\Delta_{i-1}$ and current time interval $\Delta_i$ on the heatmap shown in Figure 5. In both sub figures, we can easily see four regions that correspond to four types of transitions. The lower left region means the transition that the active states at $t_{i-1}$ remain active at $t_i$ and likewise, the upper right corner are those states remain inactive. The upper left region corresponds to inactive states changed from active states while the lower right one

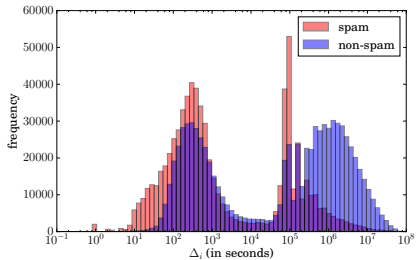

**Figure 3: Bimodal distribution of inter-arriving time for spam and non-spam reviews (Note: x-axis is in log scale)**

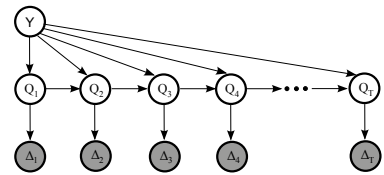

**Figure 4: Representation of Labeled Hidden Markov Model**

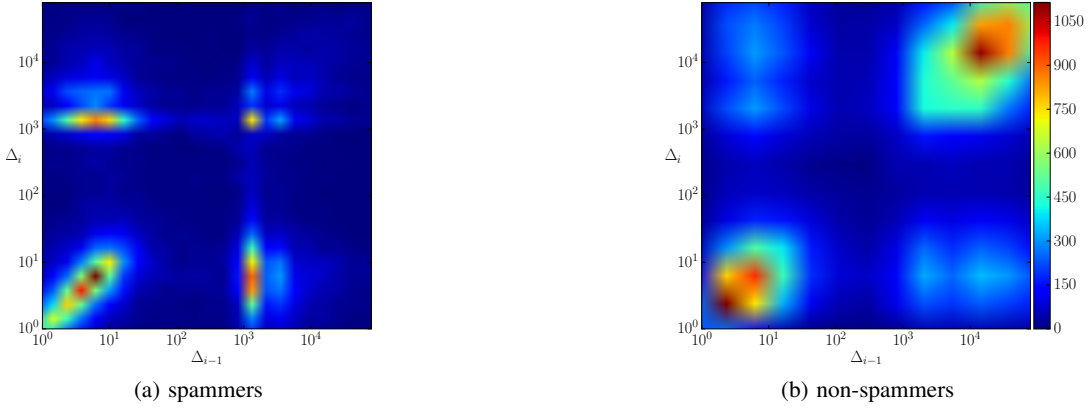

(a) spammers    (b) non-spammers

**Figure 5: Heatmap of consecutive time interval pairs (in seconds). Each point corresponds to $(\Delta_{i-1}, \Delta_i)$ for some reviewer.**

is the opposite. We find the following interesting observations:

• In the lower left corner of Figure 5(a), there is a strong positive correlation between $\Delta_{i-1}$ and $\Delta_i$ for spammers when states remain active while as the correlation between non-spammers in Figure 5(b) is very weak. This may be because even though different spammers exhibit different posting rates while in active state, the posting rates for a single spammer will not change much. But posting rates of an ordinal reviewer in active state at different timestamps may vary. As a consequence, we can see the hot area in the lower left forms a line along the diagonal for Figure 5(a) but not for Figure 5(b).

• In Figure 5(a), when spammers' states change from inactive to active (lower right region), they are active "differently" and when their states transit from active to inactive, they are inactive "similarly" (upper left region). However, it is exactly the opposite in Figure 5(b) for non-spammers, i.e., during transitions between two different states ordinary reviewers are active "similarly" but inactive "differently". Such dramatic differences suggest that spammers behave strangely at the moment of activation and hibernation and their transitions are indeed abnormal.

• Referring to the upper right corner of Figure 5(a) and Figure(b), we find that spammers "rest" in the similar rate concentrating in a small region. However, while authentic users are resting, the time intervals between reviews spread out the entire upper right corner. Clearly patterns for non-spammers are more natural or organic.

Based on the discovery of the major differences between emission probability and transition probability of HMM that ran on two classes of users, we propose a novel extension to the two-mode HMM and call it the Labeled Hidden Markov Model (LHMM) which incorporates the class label available in our dataset. The parameters of Labeled HMM are learned from the training data which is then used for prediction on the testing data using the Baum-Welch method [27]. Based on the original two-mode HMM model, we introduce a new binary variable $Y$ to represent the classes or labels as shown in Figure 4. $Y = +$ stands for spammers and $-$ for non-spammers. The variable $Y$ plays a significant role in the generating process of HMM. The transition probability matrix $\mathbb{A}$ is extended to $\mathbb{A}^+$ and $\mathbb{A}^-$ for spammers and non-spammers respectively. The set of rate parameters $<\lambda_0, \lambda_1>$ now becomes $<\lambda_0^+, \lambda_0^-, \lambda_1^+, \lambda_1^->$. Consequently, the emission probability is dependent on the user class $Y$ (13).

$$\Delta_i \sim \begin{cases} Exp(\lambda_0^Y), & Q_i = 0 \\ Exp(\lambda_1^Y), & Q_i = 1 \end{cases} \quad (13)$$

In order to predict the value of $Y$ given the observations $\Delta_{1:T}$, we need to use Bayesian theorem. The most probable value that the class variable takes is the one that better explains or generates the observations. Thus we have the following:

$$\begin{aligned} y^* &= \operatorname*{argmax}_y P(Y = y | \Delta_{1:T}) \\ &= \operatorname*{argmax}_y \frac{P(\Delta_{1:T} | Y = y) \cdot P(Y = y)}{P(\Delta_{1:T})} \end{aligned} \quad (14)$$

The denominator $P(\Delta_{1:T})$ in (14) is a constant term regardless of $y$, so we can simply drop it. The prior probability of the class variable $P(Y)$ can be easily computed by counting. The difficult part is the conditional probability $P(\Delta_{1:T}|Y)$. Recall that equation (7) is the joint probability of observations and hidden states, the conditional probability can be calculated by marginalizing the hidden states:

$$\begin{aligned} & P(\Delta_{1:T}|Y) \\ &= \sum_{Q_{1:T}} P(Q_{1:T}, \Delta_{1:T}|Y) \\ &= \sum_{Q_{1:T}} P(Q_1|Y) \prod_{i=2}^{T} P(\Delta_i|Q_i, Y) \prod_{i=2}^{T} P(Q_i|Q_{i-1}, Y) \end{aligned} \quad (15)$$

By its direct definition, the time complexity is $O(T \cdot 2^T)$. Fortunately, another dynamic programming algorithm named Forward-backward method [3, 27] can largely reduce it to linear time. Similar to Viterbi, the Forward-backward method caches intermediate results facilitating the computation.

We define a variable $\alpha_i(j|y) = P(\Delta_{1:i}, Q_i = j|Y)$ to store the joint probability of observations and $Q_i = j$ with all previous states $Q_{1:i-1}$ marginalized given $Y$. To do so, we first initialize $\alpha_1(j|y) = \pi_j^y \cdot b_j(\Delta_1|y)$, $j \in \{0, 1\}$ and then iteratively solve $\alpha_i(j|y)$, for $i = 2, \ldots, T$.

$$\alpha_i(j|y) = b_j(\Delta_i|y) \sum_{k \in \{0,1\}} \alpha_{i-1}(k) \, a_{kj}. \quad (16)$$

After that, we can get $P(\Delta_{1:T}|Y) = \sum_j \alpha_T(j|y)$ easily.

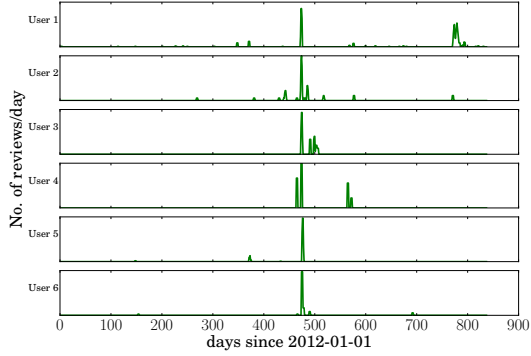

Figure 6: Examples of co-bursting behaviors

## 4. DETECTING SPAMMER GROUPS

In this section, we discuss how hidden states estimated from our model can be used to detect spammer groups. Group spamming refers to a group of reviewers writing fake reviews together to promote or to demote some target products. A spam group or a spam community is more damaging than a single individual spammer as members of a group can launch a spam attack together in a stealth mode, and due to multiple members, a group can take total control of the sentiment on a product. Each individual reviewer may not look suspicious, but a bigger picture of all of them sheds light on the collusive behaviors of a spam community. Thus, identifying such groups is very important.

Previous studies on spammer groups in [24, 34] proposed to use Frequent Itemset Mining (FIM). They treat reviewers as items and the businesses/products as transactions. Their idea is to extract groups of reviewers who have reviewed multiple products together. But it suffers from a couple of drawbacks.

- Computationally expensive: The problem is equivalent to finding all complete bi-partite subgraphs in the reviewer-product network, which is NP-hard. High support threshold in FIM will find only a few extreme cases (low recall), while low support causes combinatorial explosion especially in large datasets where there are millions of reviewers and thousands or more of products.

- Failure to capture loosely connected subgraphs: Itemsets in FIM correspond to a complete subgraph. But it is not necessarily true that every spammer should connect to all the products reviewed by other members in the same group.

- Co-reviewing doesn't mean co-spamming: There is a large chance that genuine reviewers may happen to co-review many popular products/business. Besides, nowadays recommendation systems are also suggesting consumers to buy similar products. The assumption that co-reviewing leads to co-spamming is too strong.

In our dataset, we find an interesting phenomenon that when a restaurant has bursty reviews arriving at some point, many spammers are likely to be actively writing reviews to it as well as many other restaurants. We call it co-bursting (i.e., a pair of users who have bursty reviews, some of which are posted to the same restaurant in a short period of time) as opposed to co-reviewing.

Figure 6 gives an example of six spammers' daily number of reviews. At about day 480 (since 2012/01/01), all the six users' were actively writing reviews, but they were mostly inactive at any other time periods. In addition to the temporal coincidence, we found that at least one of the reviews for each user were written for the same restaurant. Such phenomena is prevalent in our dataset showing that co-bursting is a good indicator of group spamming. Since our Hidden Markov Model gives a good estimation of hidden states for all the reviews, we thus propose to construct a co-bursting network based on the active-state of reviews. The co-burst network is more representative of the collective spamming behaviors and is thus more effective in capturing relationships between spammers than the review-product network, which were used to detect spammer groups in [24, 34]. Because it is much cleaner than reviewer-product network and the chance of random correlations is much lower. Thus it is very useful to measure the degree of collaboration between spammers.

We denote the co-bursting network as $F = \{F_{uv}\}^{n \times n}$, where $n$ is the total number of users and $F_{uv}$ represents the number of times reviewer $u$ and reviewer $v$ co-burst within a time window $\omega$ to some restaurant (rest). In our setting, we choose $\omega = 3$ days. $r_i$.state means the hidden state of review $i$ and $r_i$.t is the time when it is posted.

$$F_{uv} = \Big|(r_i, r_j) : r_i \in R^u, r_j \in R^v,\ r_i.\text{rest} = r_j.\text{rest}, \\ |r_i.\text{t} - r_j.\text{t}| < \omega,\ r_i.\text{state} = r_j.\text{state} = 1\Big| \quad (17)$$

**Algorithm 1** An efficient algorithm to construct the co-bursting network

**Input:** a set of reviews $R$, a set of users $U$, a set of restaurants $S$, time window $\omega$
**Output:** the co-bursting matrix $F$
1: **for** each $u \in U$ **do**
2: $\quad R^u = |r \in R : r.\text{user} = u|$
3: $\quad$ Run HMM on $R^u$ to get the estimated state of each of his reviews stored as $r$.state
4: **end for**
5: **for** each $s \in S$ **do**
6: $\quad R^s = |r \in R : r.\text{restaurant} = s|$
7: $\quad$ Build a B+ tree $T^s$ for $R^s$ indexing on the posting time of reviews.
8: **end for**
9: Create a hashtable $H$ to store the number of times of co-bursting for a pair of users
10: **for** each $u \in U$ **do**
11: $\quad$ **for** each $r \in R^u$ **do**
12: $\quad\quad s = r.\text{restaurant}$
13: $\quad\quad$ query B+ Tree $T^s$ to get reviews for restaurant $s$ posted between $< r.t - \omega, r.t + \omega >$ which are denoted as $C$.
14: $\quad\quad$ **for** each review $c \in C$ **do**
15: $\quad\quad\quad$ **if** $r$.state $= c$.state $= 1$ **then**
16: $\quad\quad\quad\quad i = r.\text{user},\ j = c.\text{user}$
17: $\quad\quad\quad\quad H_{i,j} = H_{i,j} + 1$
18: $\quad\quad\quad$ **end if**
19: $\quad\quad$ **end for**
20: $\quad$ **end for**
21: **end for**
22: Convert $H$ to sparse matrix $F$ and output $F$

A naive approach to construct the co-bursting network using equation 17 is very inefficient. Thus, in Algorithm 1 we propose to use a B+ tree and a hashtable to facilitate the computation. We first group reviews by its reviewer/user and run our proposed HMM model to get the estimate states for all reviews (Line 1-4) and then we build B+ tree for each restaurant to support range queries on the timestamps (Line 5-8). We maintain a hashtable to store the number of times a pair of users co-burst which is calculated efficiently from Line 10-21. The overall run-time for the last querying step is $O(m \times log(p))$ where $m$ is the total number of reviews in the dataset and $p$ is the average number of reviews written to a restaurant. Because $log(p)$ is a small constant, our proposed algorithm is linear to the size of reviews, therefore it is scalable to large datasets for any commercial review websites. Once the co-bursting networok is constructed, graph clustering can be used to find clusters, which are spammer groups (see the next section).

Table 1: Comparison of publicly available academic datasets and the Dianping dataset

| Dataset | Reviews | Business | Reviews per business | Users | Reviews per user | Users(reviews>=10) | Temporal Info | Label |
|---|---|---|---|---|---|---|---|---|
| AMT [26] | 1,600 | 20 | 80.00 | N/A | N/A | N/A | ✗ | crowdsourced |
| Yelp [25] | 66,836 | 130 | 514.13 | 34,975 | 1.92 | 690 | ✓ | pseudo label |
| Dianping | 2,762,249 | 4374 | 631.52 | 633,381 | 4.36 | 67,698 | ✓ | commercial label |

## 5. EXPERIMENTS

We now evaluate our model and compare it with baselines. Our dataset from dianping.com consists of restaurant reviews from the most popular restaurants in Shanghai, China between 2011/11/01 and 2014/04/18. As we mentioned earlier, the data has fake and non-fake labels produced by Dianping's filtering system. The labels can be trusted because Dianping has a feedback system allowing reviewers to complain. If they complain that their "genuine" reviews are removed, Dianping would send them the evidences for removing their reviews. Dianping's record shows that complaints are also quite rare. And their CTO claimed that they have used over 100 algorithms and the accuracy of their system is about 95%[4]. Dianping also has an expert group that constantly evaluates their system's performance.

Recall in section 2.1, many other researchers have compiled review datasets for opinion spam detection. We now list the statistics of all the available academic datasets that contain spam/non-spam labels in Table 1. However, due to various reasons, they are not suitable for our experiments because we focus on modeling users' temporal patterns. The Amazon Mechanical Turk (AMT) dataset is very small and contains neither users' information nor the time when reviews were posted. Even though the yelp dataset is larger, the number of reviews per user is very small because the author of this dataset only scraped 130 businesses (more precisely, restaurants) in Chicago. Besides their scraping was done per business rather than per reviewer. So in their dataset, reviews of a single reviewer do not represent the his overall behaviors because many of his reviews pertaining to other businesses are not in the dataset. There are only 690 users who wrote more than 10 reviews in the Yelp dataset. Thus, we are only able to evaluate our experiments on Dianping's dataset.

### 5.1 Performance of Labeled HMM

In our experiment, reviews are grouped by reviewers or users and sorted in the order they are posted. The parameters of Labeled HMM are learned from training data which is then used for prediction on the testing data to detect fake reviewers. We first compare our Labeled Hidden Markov Model (LHMM) with existing supervised learning methods. Although there are many recent progress on review spam, due to the lack of ground truth, most of the studies are semi-supervised or unsupervised grounded on the authors' intuitions [6, 32, 33, 35]. Since our approach is supervised learning, it is fair to compare with supervised learning models as listed below.

1. SVM(ngram) [26]: Ott et al. built a Support Vector Machines classifier using text features including unigrams and bigrams.

2. SVM(BF) [25]: Mukherjee et al. proposed many behavioral features including the number of reviews per day, rating deviation and content similarity. They showed that combining users' behaviors with linguistic features can achieve better performance.

[4] http://weibo.com/2235685314/BaoyXqlgt?type=comment

3. SVM(ngram+BF) [25]: Mukherjee et al. also combined behavioral features with ngram features improving the results.

4. PU-LEA [11]: The first Positive-Unlabeled (PU) Learning model applied in review spam detection is PU-LEA. PU learning usually outperforms traditional supervised learning when there are hidden positive instances inside negative data.

5. LHMM (UT): We would like to show how important the transitional probability of Labeled HMM (LHMM) is, so we use the uniform transition (UT) probability in LHMM rather than learning from data.

The effectiveness of all models are evaluated using the standard Accuracy, Precision, Recall and F1-score using five-fold cross validation. We can observe that although the proposed Labeled HMM (LHMM) model requires only the time-stamps of users' reviews, it markedly outperforms the baselines in review spammers detection as shown in Figure 7. It is worth noting that the largest gain of our model is recall. Because some spam accounts may exhibit mixed behaviors which confuse classifiers based on language and behavior features, whereas our proposed Labeled HMM can successfully model such temporal dynamics and therefore is most effective especially with the highest recall. Compared with LHMM(UT) which uses uniform transition probability, LHMM can achieve better results as it learns the transition probability from the data and well captures the transitional behaviors as shown in Figure 5.

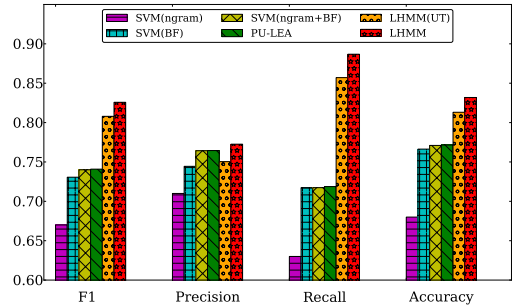

Figure 7: Model performance in Accuracy(A), Precision(P), Recall(R) and F1-score(F) (Positive class is spammer).

Based on Dianping's information, raised accounts are a popular means of spamming. Since they are raised carefully and accumulate good credits over a period of time, businesses need to pay four times more to get fake reviews from such high reputation accounts than regular fake ones in the underground market[5]. According to Dianping's commercial filters' output, over 40% of the spammers in our dataset fall into this category. Figure 8 exemplifies the daily reviews counts of three raised accounts detected by our model. Clearly, there are two distinct phases: one is the *farming phase* when the account behaves normally and randomly writes reviews to accumulate credits; the other phase is the *harvest phase* when the raised account aggressively posts spam reviews to get paid. We further investigated the effectiveness of our model in detecting raised accounts. Our proposed method successfully detected 85.41% of all the raised accounts of Dianping's dataset.

[5] http://finance.sina.com.cn/consume/puguangtai/20120904/061913036552.shtml

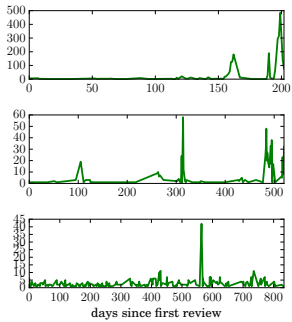

**Figure 8: Number of daily reviews of three raised accounts**

## 5.2 Performance of Co-burst Clustering

The ground truth of spammers' group affiliation is very hard, if not impossible, to know. We assume that spammers in the same cluster is a spammer group. Our goal is to validate that the co-burst graph is more intuitive and helpful in quantifying user collaborations than co-reviewing which is more noisy. We build two types of graphs: co-reviewing graph using reviewer-product relations [24, 32, 34] and co-burst graph using users' hidden states according to our definition in (17). Then we apply three efficient clustering algorithms that are suitable for the scale of our dataset (using all the data): Louvain Method [5], Kmeans and a hierarchical clustering algorithm from recent work [35].

**Table 2: Evaluation of models' performances**

| Network | Method | Purity | Entropy |
|---|---|---|---|
| Co-reviewing | Louvain | 0.69 | 0.87 |
| Co-reviewing | Kmeans | 0.72 | 0.81 |
| Co-reviewing | Hierarchical | 0.72 | 0.82 |
| Co-burst | Louvain | **0.83** | **0.67** |
| Co-burst | Kmeans | **0.86** | **0.73** |
| Co-burst | Hierarchical | **0.88** | **0.76** |

Numbers in bold indicate better performance

We use two important metrics to evaluate the clustering results: *Purity* and *Entropy*, which are widely used measures of cluster quality based on ground truth labels [2]. Purity [22] is a metric in which each cluster is assigned to the class with the majority vote in it and the accuracy of this assignment is the number of correctly assigned instances divided by the total number of instances $N$.

$$purity(C, Y) = \frac{1}{N} \sum_k max_j |y_j \cap c_k| \quad (18)$$

where $C = \{c_1, \ldots, c_k\}$ is the set of cluster ids and $Y = \{y_1, \ldots, y_j\}$ is the set of users' real class labels. $c_k$ is interpreted as the set of reviewers in cluster $k$ and $y_j$ is the set of reviewers whose label is $j$. The higher purity score means a purer cluster. Entropy [31] on the other hand measures the uniformity of a cluster. The entropy of all clusters is defined as weighted sum of entropy of each cluster:

$$entropy = -\sum_k \frac{n_k}{N} \sum_j P(j,k) \log_2 P(j,k) \quad (19)$$

where $P(j, k)$ is the probability of finding a reviewer of class $j$ in cluster $k$. The quality of a cluster improves as the entropy decreases. In Table 2, we list the purity and entropy of the clustering results. For each clustering algorithm, clusters computed from co-burst graph are significantly better than that from co-reviewing graph. Such finding is consistent with our intuition and explains the essence of spammers' collaboration.

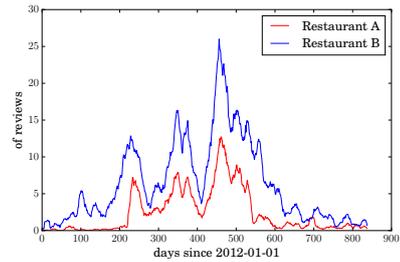

**Figure 9: Strong positive correlation between daily fake reviews of two restaurants that are only 100 meters apart**

## 5.3 Co-bursting of Restaurants Reviews

The collective spamming behaviors from spammers result in a similar view from the perspective of restaurants. Since there are many spammers actively writing reviews to a set of restaurants to promote their businesses, it is very likely to see those restaurants' time-series of daily (fake) reviews to co-burst as well. Figure 9 shows an example, which has two restaurants that are only within 100 meters. We found that there is a very strong positive correlation between their numbers of daily reviews (applied with 14-day moving average) and we notice that especially in the bursty regions, their correlation is the highest which indicates the co-bursting behaviors of restaurants. We further investigated whether they are indeed promoted by some spammer community or at least whether they were promoted by the same set of common spammers. There are overall 3196 reviewers for restaurant A and 8686 reviewers for restaurant B and interestingly they share 1166 reviewers. From April, 2013 to May, 2013 which corresponds to the highest spike of the two time-series, we found 311 reviewers wrote fake reviews to restaurant A and 591 reviewers to restaurant B and among those reviewers 139 reviewers wrote fake reviews to both restaurants. Spammer groups often proactively look for business owners to convince them to use their services. It is not surprising to see that they can help both restaurants who are competitors in the same business zone because it is easy to convince a business owner if his rival is already working with them. This explains the high correlation between their bursty regions. In summary, such views from the perspective of restaurants' bursts provide a different angle to show the intense collusion among spammer communities and explains why our model can detect hard case scenarios where traditional linguistic and behavioral features may not work.

## 6. CONCLUSION

In this work, we first conducted a series of analyses using Dianping's real-life dataset with spam labels. The analyses showed major differences of temporal patterns of spammers and non-spammers. The mean of time intervals between consecutive reviews for spammers is much smaller than that of authentic reviewers. Besides, during state transition, spammers are differently active but similarly inactive but non-spammers are exactly the opposite. We then proposed a variant of HMM named Labeled HMM to model reviewers' posting activities and to detect opinion spammers. Furthermore, we found many spammers happen to actively write fake reviews to same restaurants together in short period of time. We define a co-bursting network based on co-bursting behaviors from spammers and non-spammers. Hidden states estimated from our model are also good clues for discovering collusive spammers whose collective behaviors are well captured by co-bursting. Most importantly, since our model only leverages the time stamps of reviews, it is generally applicable to all the other review sites.